\documentclass[11pt]{article}

\usepackage[final]{acl}

\usepackage{times}
\usepackage{latexsym}

\usepackage[T1]{fontenc}
\usepackage[utf8]{inputenc}
\usepackage{graphicx}
\usepackage{microtype}
\usepackage{threeparttable}
\usepackage{inconsolata}
\usepackage[table]{xcolor}
\usepackage{xcolor}
\usepackage{arydshln}
\usepackage{graphicx}
\usepackage{amsmath}
\usepackage{booktabs}
\usepackage{multirow}
\usepackage{makecell}
\usepackage{hyperref}
\usepackage{amsfonts} 
\usepackage{booktabs} 
\usepackage{multirow}
\usepackage{xcolor} 
\usepackage{amssymb} 
\usepackage{pifont}
\usepackage{url}

\usepackage{tabularx}
\newcommand{\cmark}{\textcolor{green!70!black}{\checkmark}}
\newcommand{\xmark}{\textcolor{red!80!black}{\ding{55}}} 
\newcommand{\token}[1]{\textit{\textless #1\textgreater}}
\usepackage[most]{tcolorbox}
\newtcolorbox{promptbox}{
  colback=gray!5,
  colframe=black,
  boxrule=0.8pt,
  arc=4pt,
  left=6pt,
  right=6pt,
  top=6pt,
  bottom=6pt
}

\title{A Unified Neural Codec Language Model for Selective Editable \\ Text to Speech Generation}

\author{
 \textbf{Hanchen Pei\textsuperscript{1}},
 \textbf{Shujie Liu\textsuperscript{2}},
 \textbf{Yanqing Liu\textsuperscript{2}},
 \textbf{Jianwei Yu\textsuperscript{2}},
\\
\textbf{Yuanhang Qian\textsuperscript{1}},
 \textbf{Gongping Huang\textsuperscript{1}},
 \textbf{Sheng Zhao\textsuperscript{2}},
 \textbf{Yan Lu\textsuperscript{2}}
\\
\\
 \textsuperscript{1} School of Electronic Information, Wuhan University, China
 \\
 \textsuperscript{2} Microsoft Corporation
\\
 \small{
   \textbf{Correspondence:} \href{mailto:email@domain}{shujliu@microsoft.com}
 }
}

\begin{document}
\maketitle
\begin{abstract}
Neural codec language models achieve impressive zero-shot Text-to-Speech (TTS) by fully imitating the acoustic characteristics of a short speech prompt, including timbre, prosody, and paralinguistic information. However, such holistic imitation limits their ability to isolate and control individual attributes. In this paper, we present a unified codec language model SpeechEdit that extends zero-shot TTS with a selective control mechanism. By default, SpeechEdit reproduces the complete acoustic profile inferred from the speech prompt, but it selectively overrides only the attributes specified by explicit control instructions.
To enable controllable modeling, SpeechEdit is trained on our newly constructed LibriEdit dataset, which provides delta (difference‑aware) training pairs derived from LibriHeavy. 
Experimental results show that our approach maintains naturalness and robustness while offering flexible and localized control over desired attributes. Audio samples are available at \url{https://speech-editing.github.io/speech-editing/}.
\end{abstract}

\section{Introduction}
Recent zero-shot Text-to-Speech (TTS) generation has advanced rapidly with the rise of modern generative modeling, enabling high-fidelity voice cloning from short, unseen reference prompts. Existing systems leverage diverse acoustic representations, including discrete token-based approaches~\cite{chen2025neural,lajszczak2024base,wang2025maskgct}, continuous representations~\cite{meng2025autoregressive,eskimez2024e2,chen2025f5, wang2025felle}, and hybrid token modeling~\cite{du2024cosyvoice,yang2025pseudo,anastassiou2024seed}. Despite these advances, models treat the reference audio as a holistic, black-box condition, leaving key vocal attributes, such as timbre, emotion, prosody, and paralinguistic style, entangled and difficult to control independently.

This limitation has motivated growing interest in controllable speech synthesis, where fine-grained manipulation of attributes enables more flexible, expressive, and personalized voice generation~\cite{xie2025towards}.
Existing control paradigms include text-driven, audio-driven, and hybrid approaches. Text-driven methods rely on textual directives, including style tags~\cite{wang2025spark}, natural language descriptions~\cite{guo2023prompttts}, or instructions~\cite{zhou2024voxinstruct}, offering explicit high-level control but often failing to capture subtle acoustic details or reproduce a specific speaker’s voice. Audio-driven approaches use dual speech prompts to separately specify timbre and style~\cite{zhangvevo,zhou2025indextts2}, partially alleviating these limitations. Hybrid systems~\cite{yang2025emovoice,du2024cosyvoice} combine textual instructions with audio prompts to balance explicit control and acoustic fidelity. However, when multiple prompts are used to control different aspects of speech, interactions between them can lead to attribute leakage and conflicts, where unintended prosodic or stylistic cues affect the output. These challenges necessitate a more precise approach that supports the fine-grained editing of individual speech attributes.

In this work, we formulate controllable speech generation as a \textit{selective attribute editing problem}. Given a speech prompt $p$, a target text $x$, and an edit specification $e$, the goal is to generate speech that preserves the inherent attributes of $p$, such as speaker identity, while modifying only those explicitly indicated by $e$, such as the emotion. The editable space in this work spans three fundamental and interpretable dimensions of expressive speech:
\textit{(1)~Emotion-related attributes} describe the affective state. \textit{(2)~Prosody-related attributes} characterize paralinguistic properties such as pitch, speaking speed, and energy, which jointly determine how the utterance is realized acoustically. \textit{(3)~Speaker-related attributes} primarily correspond to timbre.
Unlike conventional TTS or global style-transfer settings, this setup requires fine-grained, attribute-level editing while keeping all unspecified components faithful to the reference. For speaker editing, the system is given an extra speaker prompt  for timbre mimic, much like voice conversion task, but within a TTS framework supporting random target text sequence.

Rather than explicitly disentangling speech attributes through specialized architectures or training schemes, we hypothesize that the in‑context learning capability of neural codec language models~(LMs)—trained on large, diverse datasets spanning multiple speakers, emotions, and vocal attributes—naturally provides implicit disentanglement.
Building on this intuition, we design SpeechEdit, which treats the speech prompt as a base canvas and selectively modifies only the attributes specified by the user. This unified formulation enables a single model to seamlessly support zero-shot TTS, voice conversion, and fine-grained style editing. 
To train SpeechEdit, we construct a new dataset, LibriEdit, by labeling the speech attributes of utterances from LibriHeavy. We introduce a Delta‑Pairs sampling method to generate training triplets (speech prompt, edit specification, speech target) by randomly sampling two utterances from LibriEdit and designating one as the prompt and the other as the target, with the differing attributes between them as the edit specification.
Experimental results conducted on various speech editing tasks show that SpeechEdit achieves highly competitive performance on naturalness and robustness, while reaching state-of-the-art~(SOTA) performance in selective speech editing.
Our main contributions are as follows:
\begin{itemize}
    \item We propose SpeechEdit, a unified selective editing framework that leverages the in‑context learning capability of neural codec LMs to integrate zero-shot TTS, voice conversion, and style editing within a single model, enabling precise attribute-level control while faithfully preserving speaker identity.
    \item We introduce a data-driven implicit disentanglement strategy that combines assumption-free Delta-Pairs sampling with our newly annotated LibirEdit dataset, enabling promising separation of speaker identity and style attributes without complex auxiliary modules and providing a scalable paradigm for expressive speech synthesis.
\end{itemize}

\begin{figure*}[t]
    \centering
    \includegraphics[width=1\linewidth]{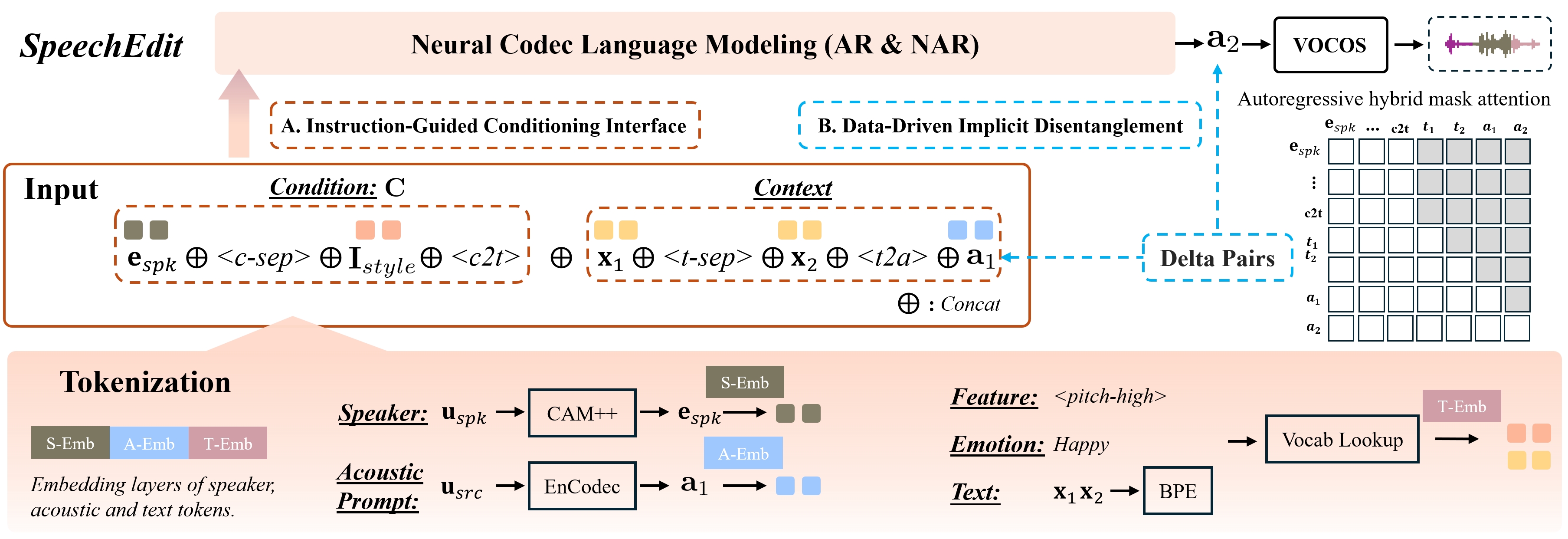}
    \caption{Overview of the SpeechEdit framework. Instruction tokens, textual content, and acoustic prompts are unified into a single token sequence through an instruction-guided conditioning interface. The codec language model performs selective attribute editing through data-driven implicit disentanglement with delta pairs.}
    \label{fig:pipeline}
    \vspace{-6pt}
\end{figure*}

\section{Related Work}
\subsection{Neural Codec LM for Speech Synthesis}
Neural codec language modeling treats speech synthesis as a sequence modeling problem over discrete acoustic tokens obtained from neural audio codecs. VALL-E~\cite{chen2025neural} pioneered this direction by proposing a hybrid Autoregressive (AR) and Non-Autoregressive (NAR) architecture. Subsequent studies have explored various aspects of neural codec LMs, including improving robustness~\cite{chen2024valle2,han2024vall,song2025ella}, efficiency~\cite{yang2025pseudo,chen2024valle2,kim2024clamtts}. Across generation architectures, codec language models involve clear trade-offs. AR models achieve strong perceptual quality by modeling temporal dependencies, but suffer from slow inference and error accumulation, while NAR and partially NAR models improve efficiency via parallel generation and duration modeling, often at the cost of temporal coherence~\cite{yang2025pseudo,wang2025maskgct}. Recent studies have explored enhancing expressive speech generation through richer conditioning signals, such as style tokens or textual instructions~\cite{ji2024textrolspeech,wang2025spark,zhou2025indextts2}, while fine-grained, attribute-level control remains challenging.

\subsection{Controllable Speech Synthesis}
Controllable speech synthesis generates natural, intelligible speech from text while enabling explicit control of specific speech attributes. Existing works explore different control dimensions, including prosody~\cite{wang2025spark}, emotion~\cite{gao2025emo}, dialect~\cite{du2024cosyvoice}, and paralinguistic features~\cite{liao2025nvspeech}. A critical challenge arises when additional controls are applied while preserving speaker identity: attribute conflict. The reference audio inherently carries its own timbre, prosody, and emotion, which can conflict with the target style specified by text or auxiliary prompts. To address this, systems typically employ either implicit or explicit disentanglement strategies. One approach, exemplified by EmoVoice~\cite{yang2025emovoice}, uses neutral reference audio to mitigate conflicts. Explicit disentanglement methods resolve conflicts through mechanisms such as gradient reversal layers~\cite{ju2024naturalspeech,zhou2025indextts2} or information bottlenecks within codebooks~\cite{zhangvevo}, which may still suffer from incomplete attribute separation and require additional model components.

\section{Proposed Method}
We formulate selective editable speech generation as a prompt-guided neural codec language modeling task, where editing is achieved by explicit instruction conditioning in the discrete codec token space. Following Encodec~\cite{defossez2022high}, a speech waveform is represented as a sequence of discrete codec tokens $\mathbf{y} \in \mathbb{Z}^{\:T \times 8}$, where $T$ is the number of time steps across $8$ codebook layers. The token $\mathbf{y}_{t,\:j}$ denotes the discrete index at time step $t$ from the $j$-th codebook layer. Building on the paradigm introduced in VALL-E~\cite{chen2025neural}, the proposed SpeechEdit extends this framework to support attribute-level speech editing via unified instruction conditioning. As shown in Figure~\ref{fig:pipeline}, both the AR and NAR stages share the same  conditioning signals. Given a speech prompt $\mathbf{a}_{1}$, its transcription $\mathbf{x}_1$, a target text $\mathbf{x}_2$, and an editing specification condition $\mathbf{C}$, the AR model predicts the first codebook layer to capture the fundamental prosodic and phonetic structure:
\begin{align}
\mathcal{L}_{\text{AR}} = - \sum_{t=1}^{T} \log p \left(\mathbf{y}_{t,1} \mid \mathbf{P}, \mathbf{y}_{<t,1}; \: \boldsymbol{\theta}_{\text{AR}} \right),
\end{align}
where $\mathbf{y}_{<t,1}$ are previously generated tokens, $\mathbf{P} = \left[ \mathbf{C}, \mathbf{x}_1, \mathbf{x}_2, \mathbf{a}_1 \right] $ is the concatenated conditioning prompt, and $\boldsymbol{\theta}_{\text{AR}}$ denotes the AR model trainable parameters. Conditioned on the first-layer predictions, the NAR model refines acoustic details by generating the subsequent layers $\mathbf{y}_{:,\:j}, j \in [2,8]$:
\begin{equation}
\mathcal{L}_{\text{NAR}} = - \sum_{t=1}^{T} \log p(\mathbf{y}_{t,\:j} \mid \mathbf{P}, \mathbf{y}_{:,\:<j}; \: \boldsymbol{\theta}_{\text{NAR}}).
\end{equation}

Unlike prior speech editing systems that rely on task-specific architectures or auxiliary disentanglement modules, SpeechEdit enables flexible and compositional attribute control through a unified instruction-driven framework.

\begin{figure*}
    \centering
    \includegraphics[width=0.85\linewidth]{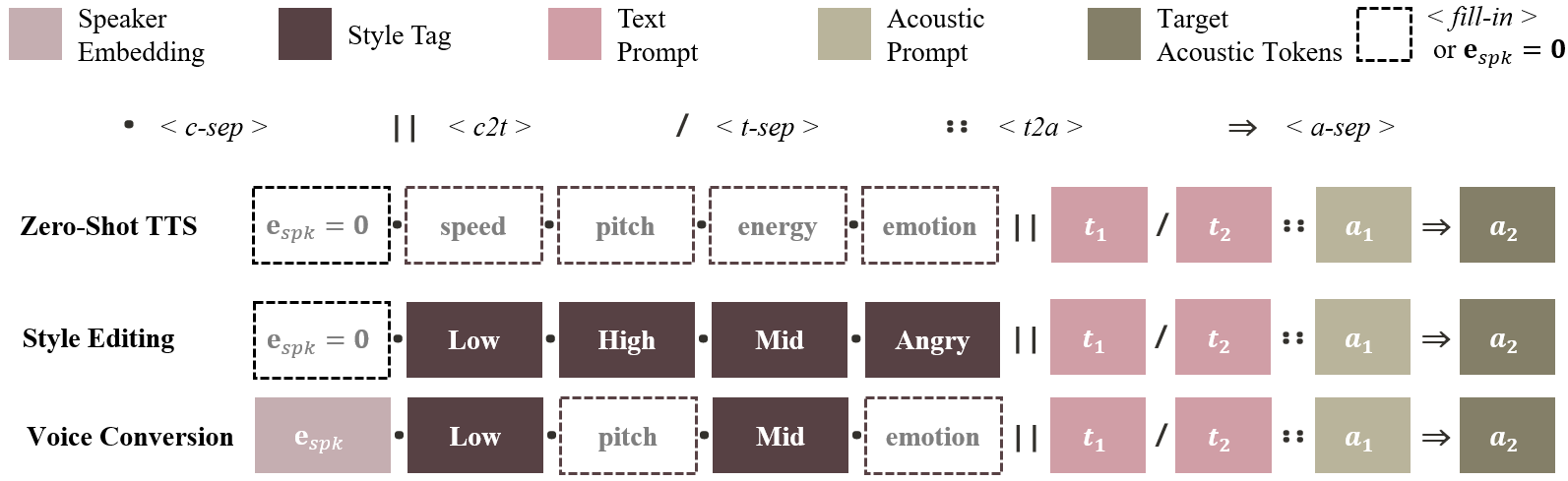}
    \caption{Token sequence composition for different tasks within SpeechEdit.}
    \label{fig:seq_in}
    \vspace{-6pt}
\end{figure*}

\subsection{Instruction Guided Interface}

We adopt a discrete instruction guided interface to model multiple speech attributes during generation as show in the bottom of Figure~\ref{fig:pipeline}.

\textbf{Categorical Attributes} Emotion and prosody attributes are represented as instruction tags. Emotion is modeled with five predefined classes: \textit{Neutral}, \textit{Happy}, \textit{Sad}, \textit{Angry}, and \textit{Surprise}. Prosody attributes, including pitch, energy, and speaking speed, are discretized into five ordinal levels ranging from \textit{Very Low} to \textit{Very High} and expressed using a structured tag format such as $\token{pitch-high}$ or $\token{speed-low}$. All instruction tags share the same vocabulary table and text embedding layer with Byte-Pair Encoding (BPE)-tokenized text.

\textbf{Speaker Identity} Speaker identity is inherently a continuous and high-dimensional factor. We extract a global speaker embedding $\mathbf{e}_{spk}$
from a speaker reference utterance using a pretrained voice print model\footnote{\url{https://github.com/alibaba-damo-academy/
3D-Speaker/tree/main/egs/3dspeaker/sv-cam++}}. This embedding is projected into the LM space via a dedicated speaker embedding layer.

\subsection{Data-Driven Implicit Disentanglement}
Instead of using auxiliary modules to decouple speech attributes, we adopt a \textit{Delta Pair Sampling} strategy to achieve data-driven implicit disentanglement, as illustrated by the blue dashed box in Figure~\ref{fig:pipeline} where training pairs are deliberately constructed with amplified attribute discrepancies to guide the model’s attention to the explicit control signals.

\textbf{Same-speaker Delta-Pair Sampling.} Two utterances from the same speaker are sampled: a source speech $\mathbf{u}_{src}$ and a target speech $\mathbf{u}_{tgt}$ from different emotion categories (e.g., $\mathbf{u}_{src}$ is Happy while $\mathbf{u}_{tgt}$ is Angry). During training, LMs are conditioned on the style tags of $\mathbf{u}_{tgt}$ but the acoustic prompt of $\mathbf{u}_{src}$. This guides the attention mechanism to: (1) extract unspecified attribute from the prompt, and (2) derive the target attributes from the style instructions.

\textbf{Cross-speaker Delta-Pair Sampling.} The source speech $\mathbf{u}_{src, spk_1}$ and target speech $\mathbf{u}_{tgt, spk_2}$ are sampled from different speakers. The model is primarily conditioned on the source acoustic tokens of $\mathbf{u}_{src, spk_1}$ along with instruction prompts for target attributes. A separate speaker reference utterance $\mathbf{u}_{ref, spk_2}$ from the target speaker provides speaker embedding to define the target identity. This reference utterance is content-independent of $\mathbf{u}_{tgt,spk_2}$, which prevents content leakage.

By conditioning the model on mismatched acoustic prompts and target instructions, the explicit instruction tokens become the only consistent signal for the attention mechanism, enabling implicit disentanglement through Delta-Pair sampling.

\subsection{Instruction Composition}
SpeechEdit unifies multiple speech generation and editing tasks within a single model by reorganizing conditioning tokens, as shown in Figure~\ref{fig:seq_in}. 
For \textit{zero-shot TTS}, setting $\mathbf{e}_{spk} = \mathbf{0}$ and all style tokens to $\token{fill-in}$ forces the model to rely entirely on the acoustic prompt for timbre and prosody. For \textit{style editing}, specific style tags are explicitly overridden. By remaining assumption‑free with respect to training‑pair attributes, our Delta‑Pair sampling strategy ensures that explicit style instructions consistently override the prompt when the two are in conflict. For \textit{voice conversion}, a target speaker embedding $\mathbf{e}_{spk}$ specifies the new identity. Style tokens can be a hybrid of explicit tags and $\token{fill-in}$, allowing prosody transfer or partial editing.
The final input sequence is structured as:
\begin{align} 
\mathbb{S}_{in} = &\underbrace{\left[ \mathbf{e}_{spk} \oplus \token{c-sep} \oplus \mathbf{I}_{style} \right]}_{\text{Conditioning}} \oplus \token{c2t} \oplus \\ \nonumber
&\underbrace{\left[ \mathbf{x}_{1} \oplus \token{t-sep} \oplus \mathbf{x}_{2} \oplus \token{t2a} \oplus \mathbf{a}_{1} \right]}_{\text{Context}},
\end{align}
where $\token{c-sep}$, $\token{t-sep}$, and $\token{a-sep}$ separate elements within the same block, while $\token{c2t}$ and $\token{t2a}$ indicate transitions across modalities, marking boundaries between global conditioning, text, and acoustic prompts.

\section{LibriEdit Dataset}
\subsection{Overview of LibriEdit}

While emotionally or stylistically expressive speech can be collected at scale and efficiently annotated using LLMs, the effective data volume is often shrinks drastically once speaker annotations are required, as shown in Table~\ref{tab:dataset-comparison}, which severely limits the ability to learn fine-grained, speaker-preserving attribute control.
To address this limitation, we build a style-labeled corpus based on the LibriHeavy dataset~\cite{kang2024libriheavy} which is chosen for three reasons:
(1) it provides a large-scale collection of read speech, with over 50k hours in its large split;
(2) audiobook narration naturally contains expressive yet non-exaggerated emotional cues that well aligned with daily speaking styles, and
(3) it offers reliable speaker identities, enabling speaker-consistent style mining.
The resulting LibriEdit dataset comprises 2566 speakers with a total 708 hours of speech. 

\begin{figure}[t]
    \centering
    \includegraphics[width=1\linewidth]{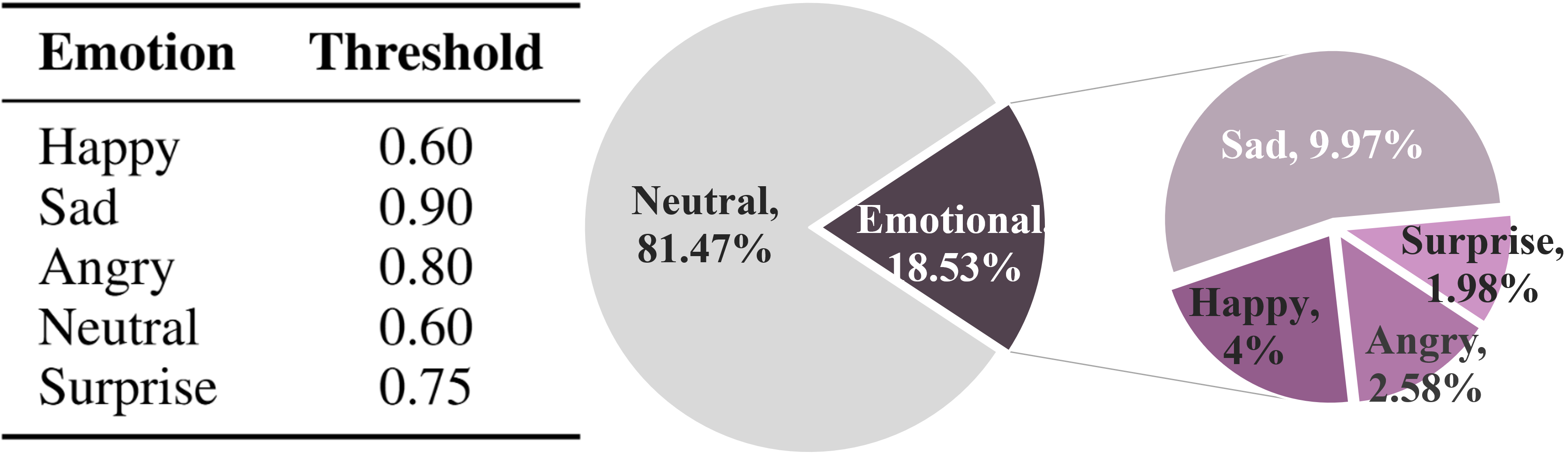}
    \caption{Confidence thresholds for emotion labeling and the resulting distribution of emotions in LibriEdit.}
    \label{fig:pie_of_emo}
    \vspace{-8pt}
\end{figure}

\subsection{Dataset Construction Pipeline}
Our LibriEdit is constructed following three steps: segmentation, emotion annotation and other attribute annotation.

\textit{Step~1: Preprocessing and Fine-Grained Sentence Segmentation.} Following the official LibriHeavy script, we begin by cutting long audiobook chapters into sentence-level segments. However, audiobook narration often exhibits style variation within a single sentence, such as neutral narration interleaved with emotionally expressive quoted speech, which remains too coarse for style labeling. So, we further refine the segmentation using Montreal Forced Aligner\footnote{\url{https://github.com/MontrealCorpusTools/Montreal-Forced-Aligner}} by splitting at breath-group boundaries and punctuation-aligned pauses. This yields shorter prosodic segments with more consistent speaking styles. A minimum duration of 2 seconds is enforced to ensure sufficient acoustic context.

\textit{Step~2: Emotion Annotation.} We begin by automatically labeling the emotion of each segment using a categorical speech emotion recognition model~(SER)~\footnote{\url{https://huggingface.co/3loi/SER-Odyssey-Baseline-WavLM-Categorical}}, which predicts an 8-way emotion distribution and outputs a confidence score for each category.
Preliminary analysis shows that the categories \emph{fear}, \emph{disgust}, and \emph{contempt} are highly ambiguous and low perceptual consistency. We therefore discard these classes and retain five reliably distinguishable emotions: \emph{neutral}, \emph{happy}, \emph{sad}, \emph{angry}, and \emph{surprise}. To improve label reliability, we apply emotion-specific confidence thresholds and keep only segments whose predicted probabilities exceed the corresponding thresholds, as summarized in the left of Figure~\ref{fig:pie_of_emo}.
We further refine the emotion labels via multi-model cross-validation with emotion2Vec-plus-large~\cite{ma2024emotion2vec} and Audio Flamingo 3~\cite{ghosh2025audio2}. A majority voting scheme is adopted, preserving only segments agreed upon by at least two models and discarding those with disagreement. The final label is corrected to the majority decision and the prompting strategy used for Audio Flamingo 3 is provided in the Appendix~\ref{sec:appendix-af}. The distribution of emotion-labeled data is shown in the right of Figure~\ref{fig:pie_of_emo}, totaling 129 hours of emotional speech.

\textit{Step~3: Prosody Attribute Annotation.} In addition to emotion labels, we annotate speed, pitch, and energy using signal-processing-based estimators.

\section{Experiment Setup}
\subsection{Implementation Detail}

\textit{Training Dataset.} We train the SpeechEdit model on the annotated LibriEdit corpus with same-speaker and cross-speaker delta pair sampling each accounting for 50\% of the data, covering diverse variations in prosody and emotional expression.

\textit{Model Configuration.} Both AR and NAR stages of SpeechEdit share a consistent backbone: a 12-layer decoder-only Transformer with 16 attention heads per layer, an embedding dimension of 1,024, and a feed-forward network with a dimensionality of 4,096 with ReLU activation. To enhance contextual modeling, the first-stage AR model employs a modified causal mask that allows bidirectional attention over prefix conditional tokens while maintaining causal attention on the following context tokens. Transcriptions are tokenized using BPE, and audio waveforms are discretized into speech tokens using the open-source EnCodec\footnote{\url{https://github.com/facebookresearch/encodec}} operating at a 6~kbps bitrate for 24~kHz audio.

\textit{Training and inference.} Both stages are trained on 16 NVIDIA Tesla V100 GPUs (32GB), with a maximum batch size of 10k tokens per GPU. The model is optimized using Adam with $\beta = (0.9, 0.98)$ and a weight decay of $0.01$. We employ an inverse square-root learning rate schedule with linear warm-up, where the learning rate increases linearly from $0$ to $5 \times 10^{-4}$ over the first 32k update steps, followed by inverse square-root decay. SpeechEdit is first pretrained on LibriHeavy-large following the VALL-E setup for 800k updates, and then further trained on the target training dataset for an additional 800k updates. The same optimization strategy is applied in both stages, with all model parameters updated. During inference, we adopt the decoding strategy of \citet{chen2024valle2}, using top-$p$ sampling with a repetition penalty.

\subsection{Baselines and Evaluation Metrics}
We compare SpeechEdit with four SOTA systems: VALL-E~\cite{chen2025neural}, which shares a similar backbone for fair zero-shot comparison; Step-Audio-EditX~\cite{yan2025step} is included as the most relevant baseline, as it is the latest open-source LM-based framework specifically optimized for unified and iterative speech editing;  CosyVoice 2~\cite{du2024cosyvoice} and IndexTTS 2~\cite{zhou2025indextts2} which are leading open-source models for instruction-based multi-style emotional synthesis.

We evaluate synthesized speech using four objective metrics:

\textbf{Word Error Rate (WER):} assesses the intelligibility by comparing the transcription of the generated audio from a Conformer-Transducer ASR model~\cite{gulati2020conformer} with ground-truth text.

\textbf{Speaker Similarity (SIM):} measures  cosine similarity between speaker embeddings extracted from the reference speech and the synthesized speech using WavLM-TDNN\footnote{\url{https://github.com/microsoft/UniSpeech/tree/main/downstreams/speaker\_verification\#pre-trained-models}}~\cite{chen2022wavlm}, indicating preservation of speaker identity.

\textbf{DNSMOS:} evaluates overall perceptual audio quality using a non-intrusive DNSMOS model~\cite{reddy2021dnsmos} trained on human ratings collected following the ITU-T P.808 protocol, with scores from 1 to 5.

\textbf{Emotion Classification Accuracy (ECA):} measures correctness of emotion expression using a WavLM-based classifier~\cite{goncalves2024odyssey}, with higher accuracy indicating stronger emotion controllability.controllability.

\begin{table*}[h]
    \centering
    \caption{Overall objective performance comparison including zero-shot TTS results on the LibriSpeech test-clean set and emotion editing results under different task settings, with the best-performing values highlighted in bold and the second-best underlined.}
    \vspace{-6pt}
    \label{tab:zero-shot-and-emotion}
    \begin{threeparttable}
    \renewcommand{\arraystretch}{0.92} 
    \setlength{\tabcolsep}{4pt}
    \small
    \begin{tabular}{llcccccc}
        \toprule
        \textbf{Task} & \textbf{Model} & \textbf{Params} & \textbf{\#~/~h} & \textbf{WER~(\%)} $\downarrow$ & \textbf{SIM} $\uparrow$  & \textbf{DNSMOS} $\uparrow$ &\textbf{ECA~(\%)} $\uparrow$  \\
        \midrule
        
        \multirow{5}{*}{\shortstack[l]{\textit{Zero-shot TTS}}} 
        & Step-Audio-EditX                          & 3~B   & -           & 1.6 & \textbf{0.63} & 3.32 & - \\
        & VALL-E-\textit{A1}$^\dagger$              & 0.5~B & 5~k         & 2.1 & \underline{0.61} & 4.00 & - \\
        & VALL-E-\textit{A2}$^\dagger$              & 0.5~B & 1~k         & 2.7 & 0.48 & \textbf{4.02} & - \\
        \rowcolor{gray!20}
        & SpeechEdit                                & 0.5~B & 0.8~k$^*$  & \textbf{1.3} & 0.48 & 4.00 & - \\
        \hdashline
        & SpeechEdit-\textit{Ablation-Data}         & 0.5~B & 0.8~k$^*$  & 1.9 & 0.45 & \underline{4.01} & - \\
        & SpeechEdit-\textit{Ablation-Task}         & 0.5~B & 0.8~k$^*$  & \underline{1.5} & 0.53 & \textbf{4.02} & - \\
        \midrule
        \multirow{9}{*}{\shortstack[l]{\textit{Emotion Easy Task}}} 
        & \textcolor{gray}{Step-Audio-EditX-$iter_0$}   & \textcolor{gray}{3~B}  & \textcolor{gray}{-} & \textcolor{gray}{1.4} & \textcolor{gray}{0.49} & \textcolor{gray}{3.39} & \textcolor{gray}{50.00} \\
        & Step-Audio-EditX-$iter_1$   & 3~B  & -                & 1.7 & 0.42 & 3.34 & 56.25 \\
        & Step-Audio-EditX-$iter_2$   & 3~B  & -                & 1.6 & 0.36 & 3.29 & 57.50 \\
        & CosyVoice 2                 &  0.5~B & $<$1.5~k$^*$      & 4.1 & \textbf{0.52} & \textbf{4.01} & 43.75\\
        & IndexTTS 2                  &  1.5~B & 135$^*$        & 2.5 & 0.44 & 3.72 & 56.25\\
        \rowcolor{gray!20}
        & SpeechEdit-\textit{C1}      & 0.5~B  & 129$^*$        & 2.5 & \underline{0.45} & \textbf{4.01} & 63.75\\
        & SpeechEdit-\textit{C2}      & 0.5~B  & 129$^*$        & 3.9 & 0.37 & 3.98 & 78.75\\
        & SpeechEdit-\textit{C3}      & 0.5~B  & 129$^*$        & 6.8 & 0.25 & \underline{4.00} & \textbf{91.25}\\
        \hdashline
        & SpeechEdit-\textit{Ablation-Data-C1}        & 0.5~B  & 129$^*$        & 3.2 & 0.40 & \underline{4.00} & 60.00\\
        & SpeechEdit-\textit{Ablation-Data-C2}        & 0.5~B  & 129$^*$        & 5.1 & 0.30 & 3.93 & 76.25\\
        & SpeechEdit-\textit{Ablation-Data-C3}        & 0.5~B  & 129$^*$        & 9.0 & 0.21 & 3.89 & \underline{82.50}\\
        
        \midrule
        \multirow{3}{*}{\shortstack[l]{\textit{Emotion Hard Task}}} 
        & CosyVoice 2                       &  0.5~B & $<$1.5~k$^*$      & 5.8 & \textbf{0.40} & \underline{3.70} & \underline{79.00}\\
        & IndexTTS 2                        &  1.5~B & 135$^*$        & \textbf{2.0} & \underline{0.39} & 3.38 & 73.00\\
        \rowcolor{gray!20}
        & SpeechEdit                        &  0.5~B & 129$^*$        & \underline{2.5} & 0.33 & \textbf{4.03} & \textbf{92.00}\\
        \hdashline
        & SpeechEdit-\textit{Ablation-Data}          &  0.5~B & 129$^*$        & 3.7 & 0.33 & 3.83 & \textbf{92.00}\\
        \bottomrule
    \end{tabular}
    \begin{tablenotes}
    \item \textbf{Params} refers to the number of parameters in the AR model. \textbf{\#~/~h} indicates the amount of training data in hours.
    \item[$*$] Indicates the amount of task-specific training data used after model initialization.
    \end{tablenotes}
    \end{threeparttable}
\end{table*}

\section{Evaluation Results}
\label{sec:results}
\subsection{Objective Evaluation}
Table~\ref{tab:zero-shot-and-emotion} summarizes the objective results, showing zero-shot TTS performance in the upper section and emotion editing in the lower section, with the best-performing values highlighted in bold and the second-best underlined.

\textbf{Zero-shot TTS.} We first evaluate zero-shot TTS performance on the LibriSpeech test clean set, with comparisons to baselines reported in Table~\ref{tab:zero-shot-and-emotion}. Results marked with $\dagger$ are cited from~\cite{chen2025neural}, focusing on ablation settings comparable training data scales. We follow the original evaluation protocol by performing five times samplings per utterance and reporting the final result by jointly ranking speaker similarity and WER. Under a restricted training budget of less than 1k hours, SpeechEdit achieves a WER of 1.3\%, outperforming VALL-E-\textit{A1} and VALL-E-\textit{A2}. Compared to Step-Audio-EditX, our model uses much less training data and fewer parameters, yet achieves substantially higher perceptual quality. Speaker similarity is slightly lower than some baselines, which is expected given the expressive prosody and diversity of LibriEdit, but overall the model maintains a strong balance between intelligibility, speaker identity, and perceptual quality under limited data.

To comprehensively evaluate the model's capability in emotion editing, we designed two experimental setups based on the relationship between the speech prompt and the target emotion:
\textit{(1)~Easy Task:} uses neutral prompts, presenting no emotional conflict with the target. It includes 80 test samples from 4 unseen speakers in the Step-audio-EditX benchmark, where the target emotions are balanced across the four non-neutral emotion categories.
\textit{(2)~Hard Task:} includes prompts with conflicting emotions in 80\% of the cases, using 100 samples from 4 unseen speakers in the LibriEdit dataset, with five target emotions roughly balanced.

\begin{figure}[h]
    \centering
    \includegraphics[width=1\linewidth]{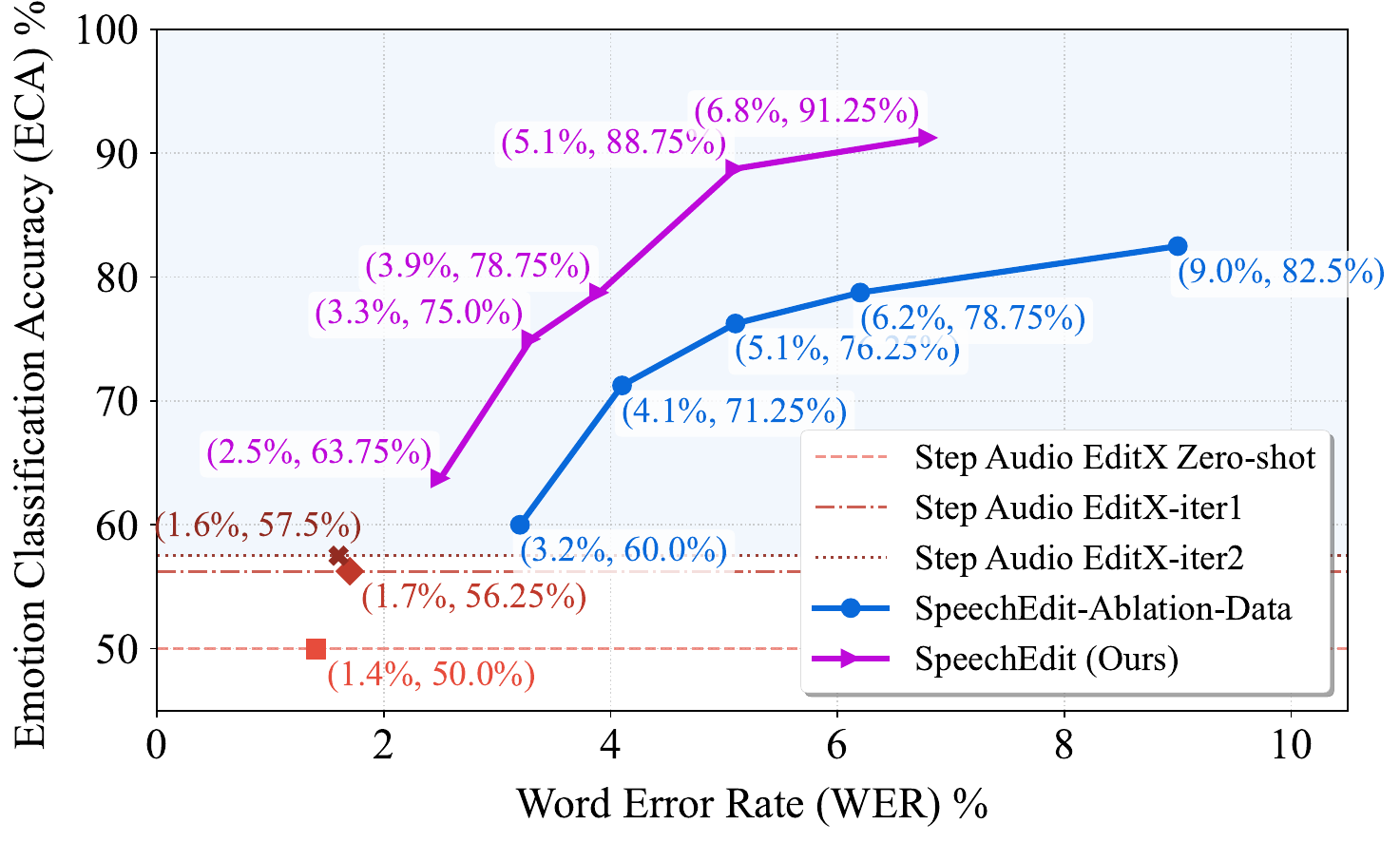}
    \caption{Emotion editing performance on the easy task.}
    \label{fig:emo_tradeoff}
\end{figure}

\begin{figure}[t]
    \centering
    \includegraphics[width=1\linewidth, trim=20 10 60 20,
    clip]{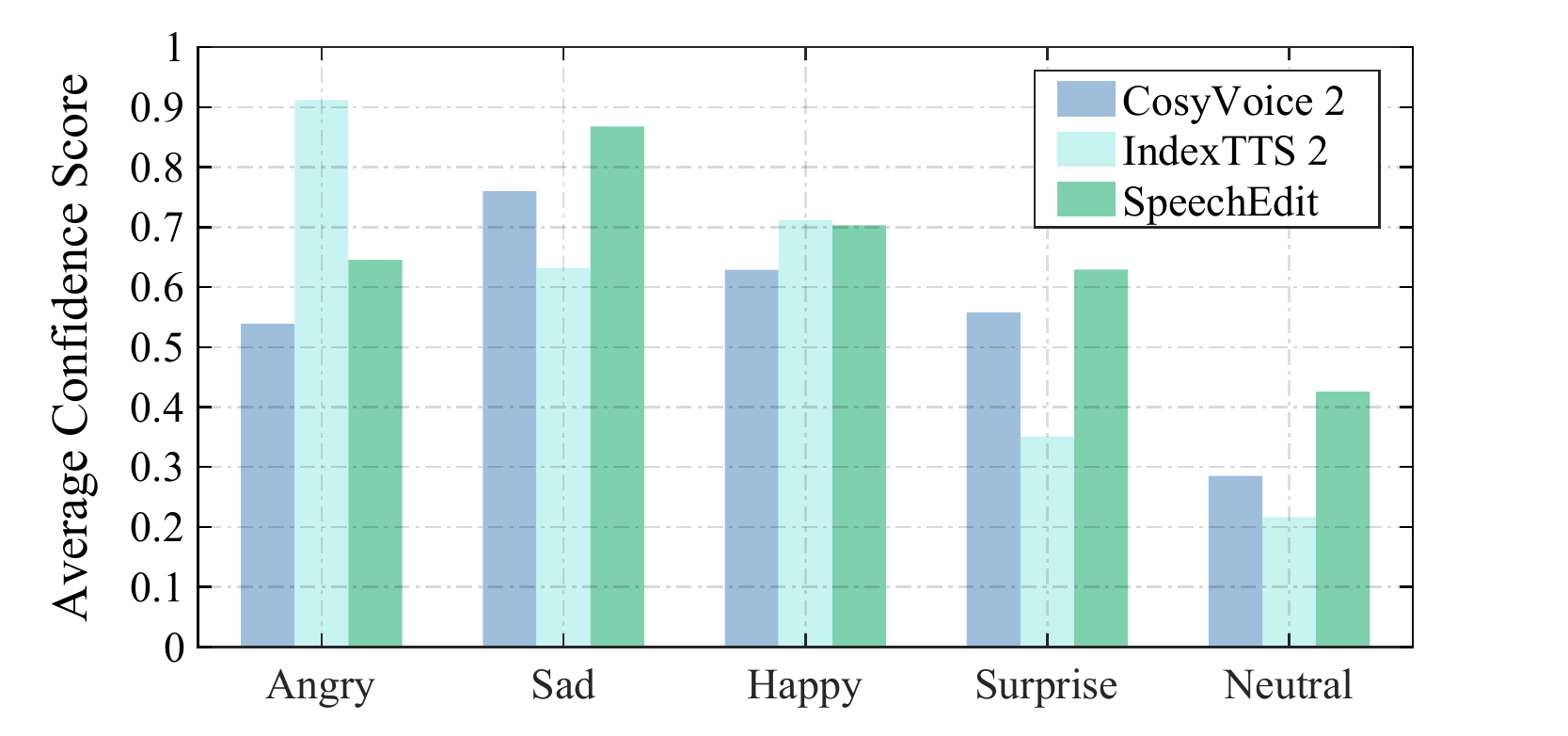}
    \caption{Average classification confidence scores for correctly predicted samples across five emotions.}
    \label{fig:mix_emo_confidence}
    \vspace{-8pt}
\end{figure}

\textbf{Emotion Edit.}
In the \textit{Easy Task}, Table~\ref{tab:zero-shot-and-emotion} reports the results for Step-Audio-EditX iterative editing and SpeechEdit, with SpeechEdit achieving the best performance across all metrics except WER. Figure~\ref{fig:emo_tradeoff} further visualizes the relationship between WER and ECA. Step-Audio-EditX does not support direct emotion-conditioned generation and instead performs zero-shot TTS ($iter_0$) followed by iterative emotion editing with the speech content fixed. Zero-shot generation achieves 50\% ECA, and iterative editing increases it only slightly to 56.25\% and 57.5\%, with negligible gain from the second iteration, indicating limited emotion controllability. In contrast, SpeechEdit performs direct emotion-controlled generation in a single stage. For each utterance, we generate five samples independently under a fixed inference configuration. Selecting only the sample with the lowest WER, SpeechEdit achieves 63.75\% ECA at an average WER of 2.5\%, already surpassing Step-Audio-EditX in emotion expression. Including samples with higher WER, ECA rises monotonically to 75\% at 3.3\% WER and 91.25\% at 6.8\% WER, illustrating stronger emotion controllability and a clear trade-off between content fidelity and emotional expressiveness, which is consistent with the fact that automatic speech recognition models tend to be less accurate on emotional or expressive speech.

In the \textit{Hard Task}, While IndexTTS 2 yields the lowest WER, SpeechEdit maintains a competitive WER of 2.5\%. SpeechEdit achieves an ECA of 92\%, substantially outperforming CosyVoice 2 (79\%) and IndexTTS 2 (73\%), indicating its ability to suppress the original emotional content from the prompt and accurately reconstruct the target emotion. It also achieves the highest DNSMOS, reflecting superior perceptual quality. Speaker similarity is slightly lower than the baselines, which is expected since SIM is computed with respect to the prompt speech. Stronger emotion modifications can alter emotion-related acoustic characteristics, naturally affecting similarity scores even when speaker identity is largely preserved. In addition, most baseline systems adopt flow-matching-based continuous-domain modeling in the second stage, which may contribute to better preservation of fine-grained acoustic details.

To further analyze emotion expression, we compute the average SER classification confidence for samples correctly generated with the target emotion. Higher confidence indicates stronger and more distinguishable emotion expression. As shown in Figure~\ref{fig:mix_emo_confidence}, SpeechEdit consistently achieves higher confidence than CosyVoice 2 across all five emotion categories. IndexTTS 2 attains particularly high confidence in the Angry category (0.91 versus SpeechEdit’s 0.64), indicating especially strong angry expression for this baseline. Notably, SpeechEdit’s confidence varies across emotions, with the highest for Sad, followed by Happy, Angry, and Surprise. This ordering aligns well with the distribution of emotions in the LibriEdit dataset, where Sad is most frequent. While this trend may partially reflect differences in perceptual salience across emotions, it also highlights the influence of data scale in emotion expressiveness, suggesting that increasing training data for underrepresented emotions, such as Angry, could further enhance emotion editing performance.

\begin{figure}[t]
    \centering
    \includegraphics[width=1\linewidth, trim=0 10 28 20,
    clip]{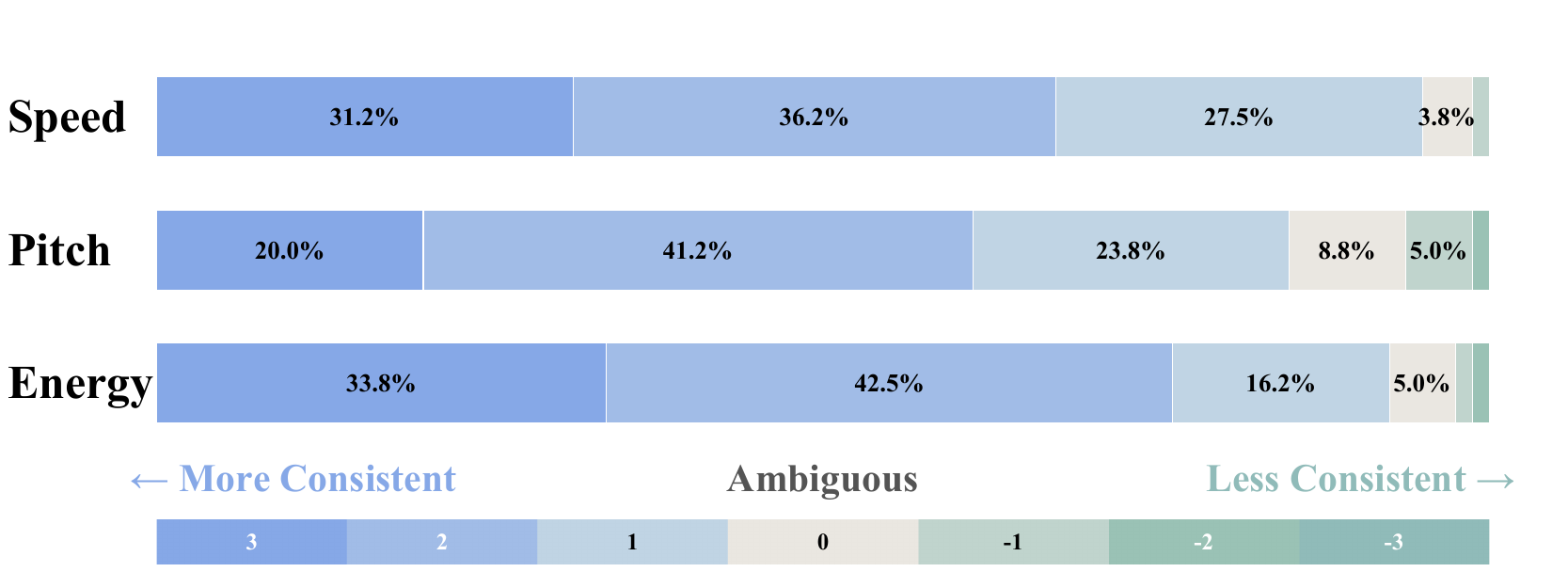}
    \caption{Result of a CMOS-style subjective test on three prosody attributes: speed, pitch, and energy.}
    \label{fig:feature_subjective}
\end{figure}
\begin{figure}[t]
    \centering
    \includegraphics[width=1\linewidth, trim=40 10 24 4,
    clip]{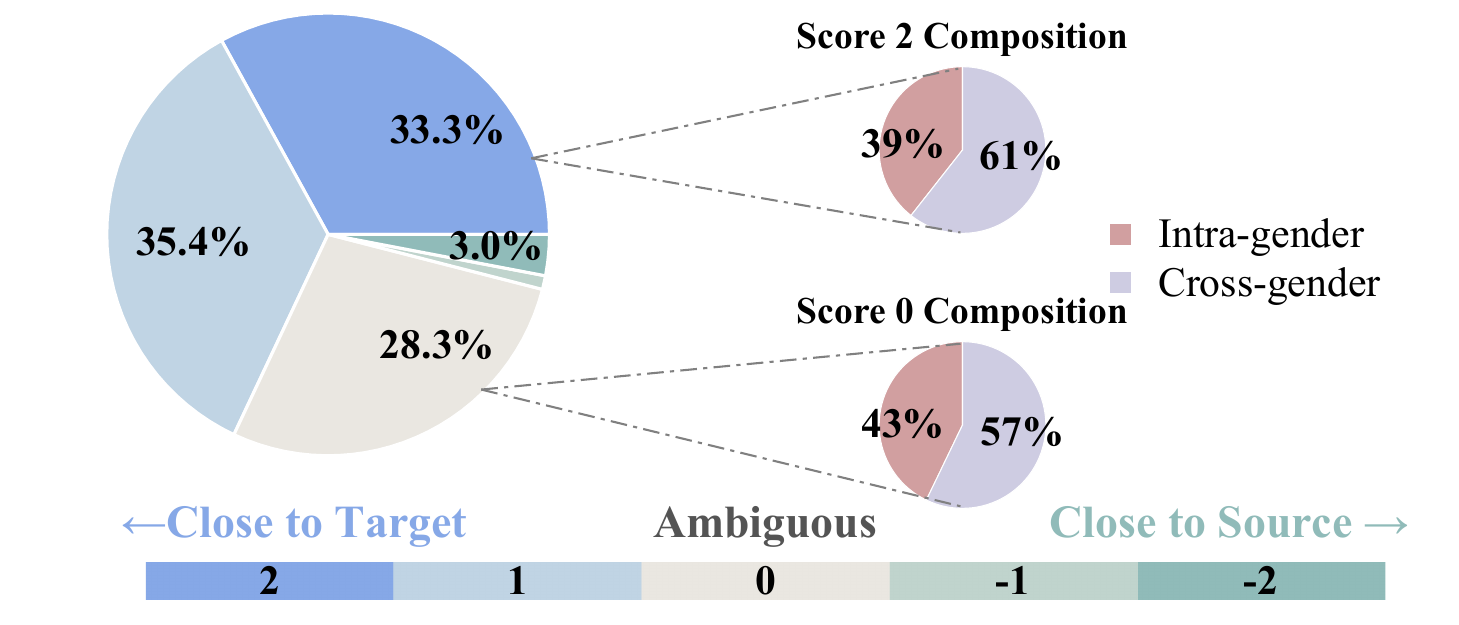}
    \caption{Result of a CMOS-style subjective test on voice conversion.}
    \label{fig:vc_subjective}
    \vspace{-4pt}
\end{figure}

\subsection{Subjective Evaluation}

We assess the model’s ability to follow style‑control instructions through a subjective test on speed, pitch, and energy.
For each test case, two speech samples are generated from the same source audio under opposite control specifications of a given prosodic attribute, such as low versus high pitch, while keeping all other factors unchanged. Ten listeners compare each pair using a comparative mean opinion score~(CMOS), where +3 indicates strong consistency with the target specification, -3 indicates clear inconsistency, and 0 denotes ambiguous perception. In addition, subjective mean opinion score~(SMOS) and subjective speaker similarity~(SSIM) are evaluated.
As shown in Figure~\ref{fig:feature_subjective}, over 85\% of the samples across all three attributes are rated consistent with the intended control direction, indicating reliable controllability. Energy control achieves the highest proportion of +3 scores~(33.8\%), followed by speed (31.2\%) and pitch (20\%). SMOS with details in the Figure~\ref{fig:smos} show that overall speech naturalness remains high, with average above 4.2. SSIM is best preserved under speed control, while pitch and energy manipulations result in slightly lower similarity and higher variance, reflecting the greater perceptual impact of these controls on speaker-related acoustic cues. We further investigate the impact of sampling temperatures on attribute controllability, with objective evaluations in \ref{appeneix-feature}

To evaluate the voice conversion capability of the proposed model, we conduct a subjective timbre similarity evaluation with four speakers, including two male and two female speakers. To ensure reliable comparison, intra-gender pairs with clearly distinct timbres are selected. A CMOS protocol is adopted, where listeners rate each sample on a five-point scale from -2 to +2. Negative scores indicate closer similarity to the source speaker, positive scores indicate closer similarity to the target speaker, and a score of 0 denotes an ambiguous identity.
Figure~\ref{fig:vc_subjective} summarizes evaluation results. Only 3.0\% of samples receive negative scores, indicating that source speaker leakage is rare. Most samples, accounting for 68.7\%, obtain positive scores, showing that the generated speech is generally perceived as closer to the target speaker. The remaining 28.3\% of samples are rated as ambiguous. When further grouped by conversion type, 61\% of samples with the highest score and 57\% of the ambiguous samples come from cross-gender conversion cases. This suggests that cross-gender conversion more readily departs from the source identity, whereas capturing fine-grained target characteristics across genders remains more difficult, sometimes leading to an intermediate timbre.

\section{Conclusion}
We presented SpeechEdit, a unified codec-LM framework for selective speech attribute editing that preserves the reference prompt’s acoustic profile while modifying only user-specified attributes. Furthermore, we constructed the LibriEdit dataset and introduced a Delta-Pairs sampling strategy to generate difference-aware training triplets, facilitating implicit disentanglement of speaker identity, prosody, and emotion without requiring specialized architectural modules. Experiments across zero-shot TTS, voice conversion, and style editing show that SpeechEdit delivers strong naturalness, robustness, and state-of-the-art selective control, suggesting that in-context learning in neural codec LMs offers a promising direction for selective and partially disentangled speech generation.

\section*{Limitations}
Despite the promising results, SpeechEdit has several limitations that warrant further investigation.
First, the granularity of speaker modeling remains a challenge. We currently employ a global speaker embedding to represent identity. While effective, this static representation may fail to capture time-varying vocal nuances or idiosyncratic articulation patterns, occasionally leading to a loss of fine-grained timbre during voice conversion.
Second, the model relies entirely on implicit disentanglement without explicit supervision. Unlike systems that employ auxiliary losses such as emotion classification or pitch regression, or use reinforcement learning to guide attribute control, SpeechEdit depends solely on in-context learning from contrastive pairs, which may limit robustness in extreme or rare attribute combinations. Third, the current controllable space is limited to emotion, prosody, and speaker identity, restricting more flexible or natural interactions, such as natural-language-based control or multi-attribute specifications. Expanding the controllable scope could enable richer and more expressive speech editing.


\bibliography{uniedit_hanchen}

\appendix

\section{Appendix}
\label{sec:appendix}

\label{sec:appendix-libriEdit}
\begin{table*}[t]
\centering
\caption{Comparison of open-sourced speech datasets in terms of fine-grained style control speech synthesis.}
\label{tab:dataset-comparison}
\small
\setlength{\tabcolsep}{3pt}
\renewcommand{\arraystretch}{0.92} 
\begin{tabular}{lccccccc}
\toprule
\multirow{2}{*}{\textbf{Dataset}} & \multirow{2}{*}{\textbf{Source}} & \multirow{2}{*}{\textbf{Speaker ID}} & \multicolumn{4}{c}{\textbf{Fine-Grained Feature Types}} & \multirow{2}{*}{\textbf{Duration~(h)}} \\
\cmidrule(lr){4-7}
& & & \textbf{Emotion} & \textbf{Speed} & \textbf{Volume} & \textbf{Pitch}  & \\
\midrule

EmoVoice-DB~\cite{yang2025emovoice}         & \multirow{1}{*}{Synthesis} & \cmark & \cmark & \xmark & \xmark & \xmark & 40 \\
\hdashline
VoxBox~\cite{wang2025spark}                 & \multirow{3}{*}{Collect}  & \xmark & \xmark & \cmark & \cmark & \xmark  & 102.5k \\
CapSpeech~\cite{wang2025capspeech}          &                            & \xmark & \cmark & \cmark & \cmark & \cmark  & 33.6k \\
TextrolSpeech~\cite{ji2024textrolspeech}    &                            & \xmark & \cmark & \cmark  & \cmark & \cmark & 300 \\
\hdashline
Expresso~\cite{nguyen2023expresso}          & \multirow{2}{*}{Record}     & \cmark & \cmark & \xmark  & \xmark & \xmark & 47 \\
EARS~\cite{richter2024ears}                 &                            & \cmark & \cmark & \cmark  & \cmark & \cmark & 60 \\
\hdashline
\textbf{LibriEdit (Ours)}                   & Audiobooks                      & \cmark & \cmark & \cmark & \cmark & \cmark & 700 \\
\bottomrule
\end{tabular}
\vspace{-4pt}
\end{table*}

\subsection{Prompt of Audio Flamingo 3}
\label{sec:appendix-af}
\begin{promptbox}
\small
\textbf{Task.} \\
Identify the emotion in the utterance. Analyze ONLY the speaker’s vocal emotion (prosody and tone), strictly ignoring the linguistic content.

\vspace{0.5em}
\textbf{Emotion Categories.} \\
Classify the emotion into one of:
\{\textit{Neutral}, \textit{Happy}, \textit{Sad}, \textit{Angry}, \textit{Surprise}\}.

\vspace{0.5em}
\textbf{Confidence Estimation.} \\
Provide a confidence score representing how certain you are about your prediction. The confidence should be a floating-point number between \textit{0.0} and \textit{1.0}, where \textit{0.0} indicates complete uncertainty and \textit{1.0} indicates complete certainty.

\vspace{0.5em}
\textbf{Output Format.} \\
Return the result \textbf{strictly} as a JSON object with two keys:
\begin{itemize}
    \item \textit{"emotion"}: the predicted emotion category,
    \item \textit{"confidence"}: the confidence score (\textit{0.0--1.0}).
\end{itemize}

Do NOT include any explanations, commentary, or extra text outside this JSON object.
\end{promptbox}

\subsection{Subjective Evaluation}
\begin{figure}[h]
    \centering
    \includegraphics[width=1\linewidth, trim = 4 10 24 20, clip]{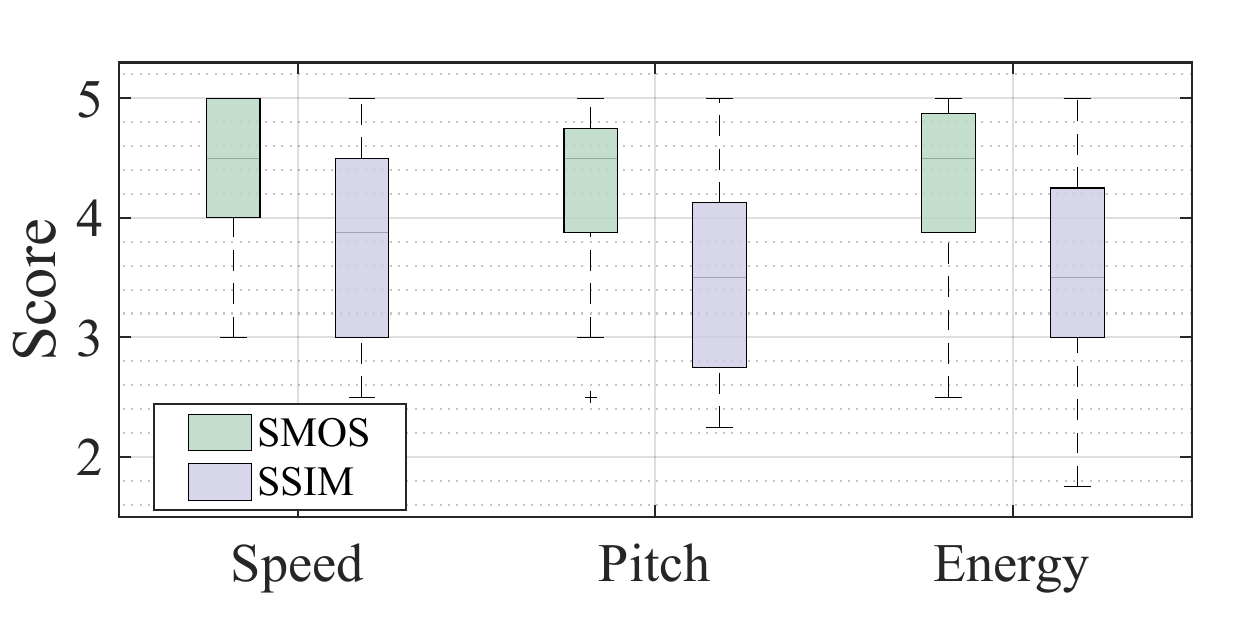}
    \caption{Score distributions of SMOS and SSIM for Speed, Pitch, and Energy.}
    \label{fig:smos}
    \vspace{-4pt}
\end{figure}

Beyond the average scores reported in the main text, Figure~\ref{fig:smos}  illustrates the detailed score distributions for three key prosodic attributes: Speed, Pitch, and Energy.
To ensure the reproducibility and consistency of our subjective testing, Table~\ref{tab:subjective_criteria} explicitly outlines the 5-point scoring criteria used for both the Subjective Mean Opinion Score (SMOS) and Subjective Speaker Similarity (SSIM).
\begin{table}[h]
    \centering
    \caption{Evaluation criteria for SMOS and SSIM.}
    \label{tab:subjective_criteria}
    \small
    \renewcommand{\arraystretch}{0.92} 
    \begin{tabularx}{\linewidth}{c c X} 
        \toprule
        \textbf{Metric} & \textbf{Score} & \textbf{Description} \\
        \midrule
        \multirow{5}{*}{\textbf{SMOS}} 
        & 5 & Excellent; natural and clear quality. \\
        & 4 & Good; minor flaws or barely perceptible noise. \\
        & 3 & Fair; perceptible degradation but intelligible. \\
        & 2 & Poor; very annoying or unpleasant to listen to. \\
        & 1 & Bad; unintelligible or totally corrupted. \\
        \midrule
        \multirow{5}{*}{\textbf{SSIM}} 
        & 5 & Identical; sounds exactly like the target speaker. \\
        & 4 & Very Similar; confident it is the same speaker. \\
        & 3 & Similar; sounds like the target but with noticeable differences. \\
        & 2 & Different; sounds like a different person. \\
        & 1 & Totally Different; no resemblance to the target speaker. \\
        \bottomrule
    \end{tabularx}
    \vspace{-8pt}
\end{table}
\subsection{Sensitivity to Sampling }
\label{appeneix-feature}
\begin{figure}[h]
    \centering
    \includegraphics[width=1\linewidth, trim = 4 4 4 4, clip]{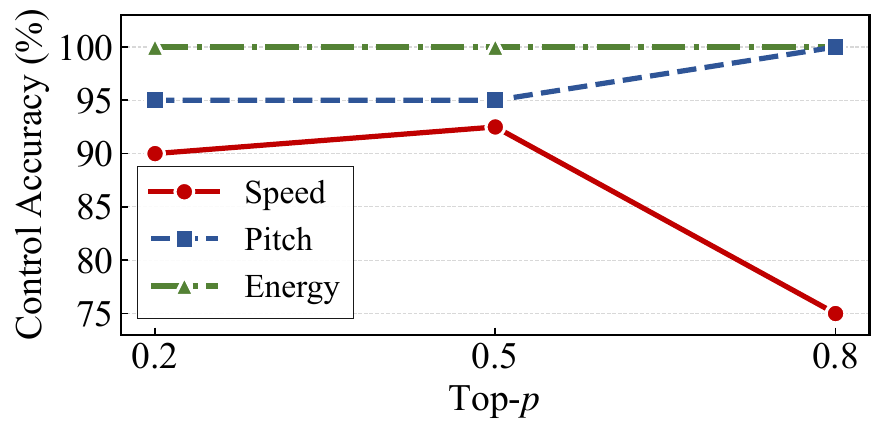}
    \caption{Impact of Top-$p$ on attribute control accuracy. }
    \label{fig:feature_acc}
    \vspace{-4pt}
\end{figure}
We investigate the impact of the sampling parameter $p$ on directional control accuracy, calculated as the percentage of samples where the attribute value follows the intended direction (i.e., $Low < High$). As shown in Figure~\ref{fig:feature_acc}, while pitch and energy exhibit strong robustness even at higher sampling rates ($p=0.8$), speed control degrades significantly to 75.0\%, indicating that excessive stochasticity compromises temporal stability. Consequently, we adopt $p=0.5$ as the optimal setting for most experiments reported in this paper, achieving the best accuracy over 92.5\%  across all attributes here. 
\subsection{Ablation Study}
We conduct ablation studies from two perspectives: (i) the training data composition and (ii) the unified task formulation.

\textbf{Data Ablation.}
Following prior works that adopt mixed training on collected emotional speech to enhance controllability, we investigate whether emotional data augmentation improves SpeechEdit, denoting this mixed-training variant as SpeechEdit-\textit{Ablation-data}. Specifically, we train SpeechEdit on a mixture of the annotated LibriEdit corpus, an internal emotional speech dataset, and the Expresso dataset~\cite{nguyen2023expresso}. The internal dataset contains approximately 30~h of acted emotional speech, while Expresso contributes an additional 5~h of professionally recorded expressive speech. In total, the training set comprises 743~h of speech. Same-speaker and cross-speaker delta pair sampling are equally balanced.

Table~\ref{tab:zero-shot-and-emotion} (below the dashed line) reports the ablation results across three tasks. Contrary to expectations, mixed training with additional emotional speech leads to consistent degradation across most objective metrics. Despite the stronger emotional expressions in the internal and Expresso datasets, Emotion-Easy task shows no improvement, while Emotion-Hard task remains comparable to the default SpeechEdit. We attribute this to data distribution mismatch and imbalance: LibriEdit contains spontaneous emotional expressions in read speech, whereas the internal and Expresso datasets comprise elicited, exaggerated emotions. This mismatch introduces a distribution shift that adversely affects training stability and weakens generalization to subtle emotional variations emphasized in Emotion-Easy. Moreover, the relatively limited scale of elicited emotional data leads to an imbalanced optimization signal, causing the model to bias toward salient emotional cues without improving fine-grained emotional controllability. These results suggest that naive emotional data augmentation via mixed training is insufficient, and that better-aligned emotional distributions and sampling strategies are essential for speech editing.

\textbf{Task Ablation.}
We observe that SpeechEdit exhibits slightly inferior speaker similarity, raising the concern that the inclusion of the voice conversion task may affect similarity preservation. To examine this effect, we train a task-ablated variant, denoted as SpeechEdit-\textit{Ablation-Task}, on the combined dataset using same-speaker delta pair sampling only, thereby removing cross-speaker supervision.
As shown in Table~\ref{tab:zero-shot-and-emotion} in zero-shot TTS, this ablated model yields a non-negligible improvement in speaker similarity from 0.45 to 0.53, indicating that the voice conversion objective introduces an inherent trade-off between identity preservation and cross-speaker controllability. While the proposed speaker-embedding-based control mechanism effectively supports voice conversion, qualitative results in Figures~\ref{fig:vc_subjective} and~\ref{fig:smos} show that the generated speech may reflect blended characteristics of the source and target speakers. This observation suggests that how to represent speaker identity in a controllable and robust manner remains an open question. More expressive and structured speaker representations may further improve conversion fidelity while preserving high speaker similarity.

\end{document}